\documentclass[aps,pra,showpacs,showkeys,amsmath,amssymb]{revtex4}
\usepackage{graphicx}% Include figure files
\usepackage{dcolumn}% Align table columns on decimal point
\usepackage{bm}% bold math
\usepackage{epsf}
\newcommand{\edc}{\end{document}}
\newcommand{\bb} {\begin{thebibliography}}
\newcommand{\eb}{\end{thebibliography}}
\newcommand{\bi}[1]{\bibitem{#1}}
\newcommand{\bc}{\begin{center}}
\newcommand{\ec}{\end{center}}

\newcommand{\be}{\begin{equation}\small}
\newcommand{\ee}{\end{equation}\normalsize}
\newcommand{\bea}{\begin{eqnarray}}
\newcommand{\eea}{\end{eqnarray}}
\newcommand{\ba}{\begin{array}{l}   }
\newcommand{\lab}[1]{\label{#1}}
\newcommand{\ea}{\end{array}}

\newcommand{\dsfrac}{\displaystyle\frac}
\newcommand{\ds} {\displaystyle}

\newcommand{\tilpsi}{\tilde{\psi}}

\newcommand{\dotpsi}{\dot{\psi}}

\def\bfr{{\bf r}}

\def\bfp{{\bf p}}
\def\al{\alpha}
\newcommand{\eps}{\epsilon}
\newcommand{\veps}{\varepsilon}

\newcommand{{\vergul}}{  ,}

\newcommand{\vecnab}{\mbox{\boldmath $\nabla$}}

\newcommand{\cale}{{\cal E}}
\newcommand{\calo}{{\cal O}}

\newcommand{\calh}{{\cal H}}
\newcommand{\calf}{{\cal F}}

\def\eps{\epsilon}
\begin{document}
%\sloppy
%\draft
%\preprint{2007-PRA}
\title{First-order quantum correction to the ground-state energy density of
 two-dimensional hard-sphere Bose atoms}
 \author{Sang-Hoon  \surname{Kim}$^{a,b}$} \email{shkim@mmu.ac.kr }
\author{Mukunda P. \surname{Das}$^{b}$}\email{mpd105@rsphysse.anu.au}
\affiliation{
$^a$ Division of Liberal Arts $\&$ Sciences, Mokpo National Maritime University,
Mokpo 530-729, Republic of Korea \\
$^b$ Department of Theoretical Physics,
RSPhysSE, Institute of Advanced Studies,
The Australian National University, Canberra, ACT 0200, Australia}
\date{\today}
%%%%%%%%%%%%%%%%%%%%%%%%%%%%%%%%%
\begin{abstract}
Divergence exponents of the first-order quantum correction of
a two-dimensional hard-sphere Bose atoms  are obtained
by an effective field theory method.
The first-order correction to the ground-state energy density
with respect to the zeroth-order is given by
 $\cale_1/\cale_0 \sim |D-2|^{-\al}|\ln\gamma|^{-\al'}$,
where $D$ is the spatial dimension, and $\gamma$ is the gas parameter ($\gamma=n a^D$).
As  $D \rightarrow 2$,  $\al =\al'=1$.
We show that the first-order quantum correction of the energy density
is not perturbative in low dimensions of $D < 2.2$
regardless of any gas parameter which is much less that 1.
\end{abstract}
\maketitle

%%%%%%%%%%%%%%%%%%%%%%%%%%%%%%%%%%%%
\section{Introduction}
%%%%%%%%%%%%%%%%%%%%%%%%%%%%%%%%%%%%

Weakly interacting uniform Bose gas is a fundamental topic
of many-body Bose systems and has been studied for more than five decades.
Bose-Einstein condensation(BEC) is known to be the only phase transition
that does not require interaction.
However, in the real system an interaction exists even in very dilute Bose gases,
and often the diluteness is explained
by a D-dimensional gas parameter: $\gamma(D)=n a^D \ll 1$,
where $n$ is the D-dimensional density and $a$ is the s-wave scattering length.
At zero temperature the chemical potential, energy density, number density,
and speed of sound are written
as an expansion in powers of the gas parameter $\gamma$.

A fundamental approximation method of a weakly interacting Bose system
was introduced by Bogoliubov in 1947 \cite{bogoliubov}.
In three-dimensions(3D) the ground-state energy density is given by
\be
\cale_0 + \cale_1 + \cdots
= \cale_0 \left[ 1 + c_1 \sqrt{\gamma} + c_2 \gamma\ln \gamma +  \cdots \right].
\label{1}
\ee
Similarly, the  number density in 3D is given by
\be
n_0 + n_1 + \cdots
= n_0 \left[ 1 + \frac{8}{3\sqrt{\pi}} \sqrt{\gamma} + \cdots \right].
\label{2}
\ee
$\cale_0 = 2 \pi \hbar^2 a n^2/m$ is the zeroth-order or mean-field  energy density.
The first-order quantum correction $c_1=128/15\sqrt{\pi}$  was first  obtained by Lee and  Yang in 1957 for a hard-sphere Bose gas \cite{lee}.
Later, it has been shown by Brueckner {\it et al.} \cite{brueckner,beliaev}
that above results are generally true for
any short-range potential with scattering length $a$.
The second-order quantum correction $c_2=8(4\pi-3\sqrt{3})/3\sqrt{\pi}$
that was originated from a three-body interaction was obtained by  Wu {\it et al.}
in 1959 \cite{wu,hugenholtz,sawada}.
Recently, an effective field theory(EFT) method  has been suggested
to obtain the coefficients(Braaten and Nieto \cite{braaten}).
This method was introduced by Georgi {\it et al.} in 1990s \cite{georgi,kaplan,manohar}.

In two-dimensions(2D) an uniform Bose gas is governed by strong long-range fluctuations.
The fluctuations inhibit the formation of a true long-range order.
Therefore,  2D uniform  Bose gas does not undergo BEC transition
at finite temperatures \cite{ketterle}.
However, this 2D system turns superfluid below a certain temperature
while under confinement.
The ground-state energy density  of hard-sphere bosons in 2D
has been studied by Schick in 1971 \cite{schick}.
\be
\cale_0 + \cale_1 + \cdots =
\cale_0 \left[ 1 + {\calo}(1/\ln \gamma )  \right],
\label{3}
\ee
and  number density is
\be
n_0 + n_1 + \cdots
= n_0 \left[ 1 + \frac{1}{\ln(1/\gamma)}  + \cdots \right].
\label{4}
\ee
$\cale_0 = 2 \pi \hbar^2 n^2/ (m |\ln \gamma |)$ is the mean-field  energy density.
The first-order  number density in 2D is finite as
$n_1/n_0 = 1/\ln(1/\gamma) \ll 1$, but the first-order quantum correction of the
ground-state energy density has not been obtained yet
because an infrared divergence of an integral prevents a direct calculation.
It has the divergence form of
\bea
\left| \frac{\cale_1}{\cale_0}  \right| &\sim& \frac{1}{|D-2|^{\al} |\ln \gamma|^{\al'}},
\label{5}
\eea
where the exponents $\al$ and $\al'$  are positive constants.

In this paper, we will calculate the two exponents of divergence
$\al$ and $\al'$ analytically.
The EFT has an advantage that it works even in non-integer dimensions.
Therefore, we can rewrite the EFT  in non-integer dimensions between two and three.
Then, taking the limit of
$D=\lim_{\eps \rightarrow o^{+} }2(1+\eps)$, we will obtain the two
exponents of divergence  $\al$ and $\al'$.
Furthermore, since the first-order quantum correction
of the ground-state energy density
 converges in D=3 but diverges  in D=2, we may find the dimensional border of the perturbative method which is effective between two and three.

The paper is organized as follows:
In Section II, we review the basic structure of EFT of a dilute Bose system in non-integer dimensions between two and three \cite{braaten,andersen,rakimov}.
In Section III, we reproduce the zeroth-order or mean-field results in D-dimensions.
In section VI, we obtain the first-order quantum correction in 2D and two exponents $\al$ and $\al'$.
Also we interpolate the first-order quantum depletion between 2D and 3D.
Finally, in section V we summarize the results.

%%%%%%%%%%%%%%%%%%%%%%%%%%%%%%%%%%%%%%%%%%%%%
\section{Effective field theory by an effective Lagrangian }
%%%%%%%%%%%%%%%%%%%%%%%%%%%%%%%%%%%%%%%%%%%%%%%%%%%%%%%%%

EFT is a general approach that can be used to analyze the low energy
behavior of a physical system
and takes advantage of the separation of scales  to make
model-independent predictions \cite{georgi,kaplan,manohar}.
The effective lagrangian that describes the low energy physics
is written in terms of only the long-wavelength degrees of freedom
that includes every non-renormalizable interactions,
but to a certain order in a low energy expansion, only a finite number of operators
contribute to a physical quantity \cite{andersen}.

An uniform Bose gas can be described by field theoretic method with Hamiltonian
${\calh}-\mu N$, where $\mu$ is the chemical potential and $N$ is the number  operator.
The number of atoms are conserved by the phase transformation: $\psi(\bfr,t)
\rightarrow e^{i\theta}\psi(\bfr,t)$.
The  free energy  in ground-state of the system is
\be
\calf(\mu)=\langle {\calh}-\mu N \rangle_{\mu},
\label{202}
\ee
where
$\langle \cdots  \rangle_{\mu}$ denotes the expectation value in the ground state.
Therefore, $\langle \calh \rangle_{\mu}=\cale(\mu)$ and $\langle N \rangle_{\mu}=n(\mu)$ are the energy  and number  in the ground state.
The free energy $\calf$ is given by all connected vacuum diagrams
that are Feynman diagrams with no external legs \cite{braaten}.
The number density $n$ is given by
the expectation value  of $ ( \psi^{\dagger} \psi )$
 in the ground-state (we set  $\hbar=1$ for convenience)
\begin{eqnarray}
n(\mu)
&=& \frac{1}{Z} \int {\cal{D} \psi^{\dagger} \cal{D} \psi  (\psi^{\dagger} \psi  ) }\,
 e^{ i S[\psi^{\dagger},\psi]}
\label{55}
\end{eqnarray}
 where $\psi^{\dagger}$ and $\psi$ are complex field operators of bosons,
  $S$ is the action from the Lagrangian  \cite{braaten},
   and $Z$ is the grand-canonical partition function   given by
\be
Z= \int {\cal{D} \psi^{\dagger} \cal{D} \psi }\, e^{i S[\psi^{\dagger},\psi] }.
 \lab{53}
\ee

The grand-canonical ensemble of uniform Bose particles
with a  two-body interaction is governed by the  action
\be
S[\psi^{\dagger},\psi]=\int dt \int d^D x
\left[ \psi^{\dagger} \left( i\frac{\partial }{\partial t}+ \frac{\vecnab^2}{2m} + \mu
\right) \psi - \frac{1}{2}g \left( \psi^{\dagger} \psi \right)^2 \right],
\lab{21}
\ee
where  $m$ is the atomic mass, and
 $g$ is the D-dimensional coupling constant
 which  contains pairwise interaction between atoms.
In D-dimensions $g$ is given by \cite{nogueira}
 \be
 g(D)=\frac{4 \pi^{D/2} a^{D-2}}{m}\frac{1}{2^{2-D}\Gamma(1-\frac{D}{2}) \gamma^{D/2-1} + \Gamma(\frac{D}{2}-1)},
\label{23}
 \ee
 where $\Gamma$ is the Gamma function.
It satisfies the two limiting cases. In 3D, it has the well-know form
\be
g(3) = \frac{4 \pi a}{m}.
\label{24} \\
\ee
In 2D, it has the logarithmic form \cite{kim,lieb}
\be
g(2)=
%-\frac{2\pi}{m}\frac{1}{ \gamma_E +\ln\sqrt{\gamma}-\ln 2 } \approx
\frac{4\pi}{m}\frac{1}{\ln(1/\gamma)}.
\label{25}
\ee

When the temperature of a Bose system falls  below the condensation temperature $T_c$,
we can write the quantum field $\psi$ in terms of a time-independent
condensate $\phi$ and a quantum fluctuation field $\tilde{\psi}$
 \be
 \psi = \phi + \tilde{\psi}.
 \lab{27}
 \ee
The fluctuation field $\tilpsi$ can be conveniently written in terms of two real fields $\psi_1$ and $\psi_2$
\be
\tilpsi = \frac{\psi_1 + i \psi_2}{\sqrt{2}}.
\label{29}
\ee
 In  the uniform system $\phi$ is the condensate order parameter and a real constant.
 It corresponds to a breaking of the global $U(1)$ symmetry
and takes into account the Bogoluibov  shift of the field operator.

   The conservation of particle numbers
 requires that $\tilpsi$  has non-zero momentum component so that
$ \langle \psi \rangle =  \phi$ and $ \langle \tilpsi \rangle=0$.
Therefore, the condensate order parameter $\phi$ defines the density of condensed particles while $\tilpsi$ defines the density of uncondensed particles.
 Then, the zeroth- and first-order quantum correction to the density  are given by
 \be
n_0=\phi^2, \quad
n_1=\langle {\tilde{\psi}}^{\dagger} \tilde{\psi} \rangle.
\lab{eq7}
 \ee

Substituting Eqs. (\ref{27}) and (\ref{29}) into Eq. (\ref{21}) of the action $S$,
we decompose the real part of the action into three parts \cite{braaten,andersen,rakimov}:
\be
S[\phi,\psi_1,\psi_2]=S_{clas}[\phi]+S_{free}[\phi,\psi_1,\psi_2]
+S_{int}[\phi,\psi_1,\psi_2].
\label{31}
\ee
$S_{clas}$ is the classical part of the action. It does not contain any filed operator:
\be
S_{clas}[\phi]=\int dt \int d^D x \left( \mu \phi^2 - \frac{1}{2}g \phi^4 \right).
\label{33}
\ee
$S_{free}$ is the free part of the action. It is quadratic in $\psi_1$ and $\psi_2$:
\be
S_{free}[\phi,\psi_1,\psi_2]=\int dt \int d^D x \left[
 \frac{1}{2}(\dotpsi_1 \psi_2 - \psi_1 \dotpsi_2 )
+ \frac{1}{2} \psi_1 \left(\frac{\vecnab^2}{2m}+X \right) \psi_1
+\frac{1}{2} \psi_2 \left( \frac{\vecnab^2}{2m}+Y \right) \psi_2\right],
\label{35}
\ee
where $\dotpsi \equiv \partial \psi / \partial t$, and the two new variable $X$ and $Y$ are
\begin{eqnarray}
X&=&\mu - 3 g \phi^2,
\label{37} \\
Y&=& \mu - g \phi^2.
\label{38}
\end{eqnarray}
The terms $3 g \phi^2$ and $g \phi^2$ in the $X$ and $Y$ are
mean-field self-energies of the system.

$S_{int}$ is the interaction part of the action. It is the remaining terms:
\be
S_{int}[\phi,\psi_1,\psi_2]=\int dt \int d^D x \left[
\sqrt{2} J \psi_1 + \frac{K}{\sqrt{2}} \psi_1 (\psi_1^2 + \psi_2^2)
-\frac{g}{8} (\psi_1^2 + \psi_2^2)^2 \right],
\label{41}
\ee
where
\be
J= \phi (\mu -  g \phi^2)=\phi Y,
\label{42}
\ee
and
\be
K= - g \phi.
\label{43}
\ee

The free part of the action in Eq. (\ref{35}) gives rise to
a propagator, which can be used in perturbative framework.
If we take the Fourier Transform in a momentum space, the fluctuating part is given by
 \be
 \tilde{\psi}(t,\bfr)=\dsfrac{1}{\sqrt{  V
}}\sum_{n=-\infty}^{\infty}\sum_{p}\tilde{\psi}(\omega_n,\bfp)
\exp\{ - i\omega_n t+i\bfp\cdot\bfr\},
 \lab{47}
 \ee
  where
$V^{-1}\ds{\sum_{p}}=\int d^Dp/(2\pi)^D$, and
$\omega_n=2\pi nT$ is the  Matsubara frequency.
Therefore, the propagator of the free action $S_{free}$  is written as
\begin{eqnarray}
G(\omega,p)=\dsfrac{1}{\omega^{2}-E_p^2 + i E_p}\left(
\begin{array}{cc}
\frac{p^2}{2m} - Y  & -i \omega \\
\nonumber
 i \omega & \frac{p^2}{2m} - X
\end{array}\right),\\
\label{eq22}
\end{eqnarray}
with the dispersion relation
\be
E_p= \sqrt{ \left(\frac{p^2}{2m} - X\right)\left(\frac{p^2}{2m} - Y\right)}.
\label{49}
\ee
This is a general form of the dispersion and includes
every information of the energy spectrum in  the two self-terms $X$ and $Y$.
It is possible to diagonalize the propagator matrix in Eq. (\ref{eq22})
by a field redefinition or renormalization,
which is equivalent to the Bogoliubov transformation in the operator method.
However, such a redefinition makes the
interaction terms in the action more complicated and increases the number of diagrams.
Therefore, we use the propagator matrix with off-diagonal  elements
to minimize the number of diagrams.

%%%%%%%%%%%%%%%%%%%%%%%%%%%%%%%%%%%%%%%%%%%%%
\section{Thermodynamic potential and zeroth-order results}
%%%%%%%%%%%%%%%%%%%%%%%%%%%%%%%%%%%%%%%%%%%%%%

It is convenient to introduce the thermodynamic potential $\Omega(\mu,\phi)$.
The thermodynamic potential contains the information required to determine
all of the thermodynamic functions.
The free energy $\calf(\mu)$ can be obtained by evaluating $\Omega$
at a particular value of $\phi$.

The sum of the vacuum graphs is independent of
the arbitrary background condensate $\phi$.
Thus, the sum of connected vacuum diagram reduces to the sum of
one-particle irreducible vacuum diagrams \cite{braaten,andersen,rakimov}:
\be
\calf(\mu) = \Omega(\mu,\phi(\mu))
\label{59}
\ee
From Eq. (\ref{202}) the ground-state energy density is written by
 \be
 \cale(\mu) = \calf(\mu) + n \mu.
 \label{57}
\ee
By differentiating the free energy, we obtain the density and chemical potential, too.
\bea
n(\mu) = -\frac{d \calf}{d \mu}, \quad
 \mu(n) = \frac{d \cale}{d n}.
 \label{204}
\eea

The $n$-loop contribution to the $\Omega$ is denoted by $\Omega_n$ in FIG. 1.
It is given by all one-particle irreducible vacuum diagrams
and can be expanded in the number of loops:
\be
\Omega(\mu,\phi)=\Omega_0(\mu,\phi) + \Omega_1(\mu,\phi) + \Omega_2(\mu,\phi)+ \cdots.
\label{61}
\ee
If $\Omega$ is evaluated at a value of the condensate,
all one-particle reducible diagrams vanish.
Then, the free energy density in Eq. (\ref{59}) is
\be
\calf(\mu) = \Omega_0(\mu,\phi) + \Omega_1(\mu,\phi) + \Omega_2(\mu,\phi)+ \cdots.
\label{63}
\ee

%%%%%%%%%%%% FIGURE 1b    %%%%%%%%%%%%%%%%%%%%%%%%%%%%%%%
\begin{figure}
\resizebox{!}{0.07\textheight}
{\includegraphics{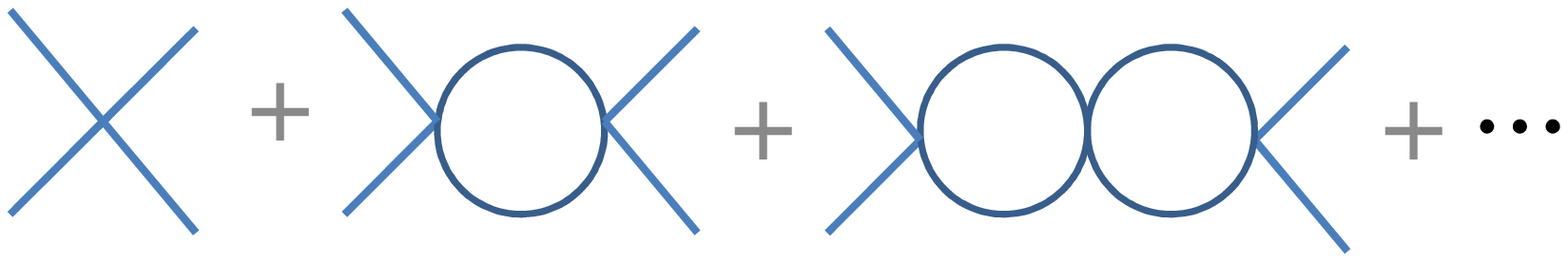}}
\caption{\label{fig:loop}
Loop diagram of the two-body interaction.
 The zeroth, first, and second loop from the left to right}
\end{figure}
%%%%%%%%%%%%%%%%%%%%%%%%%%%%%%%%%%%%%%%%%%%%%%%%%%%%%%%%%%%%

Using Eqs. (\ref{27}) and (\ref{29}), the condition of the condensate
%in Eq. (\ref{58})
reduces to $\langle \psi_1 \rangle = \langle \psi_2 \rangle =0$.
This condition is equivalent to
\be
\frac{\partial \Omega(\mu,\phi)}{\partial \phi}=
\frac{\partial \Omega_0(\mu,\phi)}{\partial \phi}
+\frac{\partial \Omega_1(\mu,\phi)}{\partial \phi}
+\frac{\partial \Omega_2(\mu,\phi)}{\partial \phi}+ \cdots
=0
\label{65}
\ee
The number density in Eq. (\ref{204})
is obtained from the thermodynamic potential, too.
\be
n(\mu) = -\frac{\partial \Omega}{\partial \mu}.
\label{66}
\ee
The free energy  can also be expanded in powers of quantum corrections around
the mean-field value $\calf_0(\mu)$:
\be
\calf(\mu) = \calf_0(\mu)  + \calf_1(\mu)  + \calf_2(\mu) + \cdots.
\label{67}
\ee

The loop expansion in Eq. (\ref{63}) does not coincide with the above expansion
of $\calf(\mu)$ in powers of quantum corrections because of its independence of $\phi$.
To obtain the expansion of $\calf$ in powers of quantum corrections, we must
expand the condensate $\phi$ around its classical minimum $\phi_0$,
which satisfies
\be
\frac{\partial \Omega_0(\mu,\phi_0)}{\partial \psi}=0.
\label{69}
\ee
By expanding Eq. (\ref{65}) in powers of $\phi-\phi_0$, and solving
for $\phi$, we obtain the quantum expansion for the condensate:
\be
\phi=\phi_{0}+\phi_{1}+\phi_{2}+\cdots,
\label{71}
\ee
where $\phi_{n}$ is the $n$th-order quantum correction.
For instance, the first-order quantum correction for the condensate $\phi_1$
is obtained from the expanding of Eq. (\ref{69}) around $\phi_0$.
The nonzero terms are
\be
\frac{\partial \Omega_0(\mu,\phi)}{\partial \phi}
=\frac{\partial^2 \Omega_0(\mu,\phi_0)}{\partial \phi^2}(\phi-\phi_0)
+ \frac{\partial \Omega_1 (\mu,\phi_0)}{\partial \phi}+\cdots=0.
\label{72}
\ee
Then, $\phi_1$ in Eq. (\ref{71}) is written as
 \be
%\vbar_1=-\frac{\frac{\partial \Omega_1 (\mu,\vbar_0)}{\partial v}}
%{\frac{\partial^2 \Omega_0 (\mu,\vbar_0)}{\partial v^2}}.
\phi_1=-\left[ \frac{\partial \Omega_1 (\mu,\phi_0)}{\partial \phi} \right]
\left[ \frac{\partial^2 \Omega_0 (\mu,\phi_0)}{\partial \phi^2} \right]^{-1}.
 \label{73}
 \ee

 Therefore, keeping up to the terms of the second order and $\phi-\phi_0 \approx \phi_1$,
 the first three terms of the free energy density is
  \begin{eqnarray}
  \calf_0(\mu)&=&\Omega_0(\mu,\phi_0),
 \label{75} \\
   \calf_1(\mu)&=&\Omega_1(\mu,\phi_0),
   \label{77} \\
 \calf_2(\mu)&=&\Omega_2(\mu,\phi_0)+\frac{\partial \Omega_1(\mu,\phi_0)}{\partial \phi}\phi_1+\frac{1}{2}\frac{\partial^2\Omega_0(\mu,\phi_0)}{\partial \phi^2} \phi_1^2.
      \label{79}
 \end{eqnarray}

The mean-field thermodynamic potential is given by the terms in the classical action in Eq. (\ref{33}) as
\be
\Omega_0(\mu,\phi)=-\frac{1}{V}S_{clas}=-\mu \phi^2 + \frac{1}{2}g  \phi^4.
\label{81}
\ee
From Eq. (\ref{69}) we obtain the classical minimum $\phi_0$ as
\be
\phi_0=\sqrt{\frac{\mu}{g}}
\label{83}
\ee
Therefore,  the mean-field free energy density in D-dimensions is obtained
from Eq. (\ref{75})
\be
  \calf_0(\mu)=\Omega_0(\mu,\phi_0)=-\frac{\mu^2}{2g}.
\label{85}
\ee
The mean-field number density is obtained from Eq. (\ref{55}) as
\be
n(\mu_0)=-\frac{\partial \calf(\mu_0)}{\partial \mu}=\frac{\mu_0}{g}
\label{87}
\ee

The mean-field ground-state energy density in D-dimensions
is obtained from Eq. (\ref{57})
\be
\cale_0^{(D)} = -\frac{\mu_0^2}{2g} + n\mu_0 = \frac{1}{2}g(D) n^2 .
\label{88}
\ee
In 3D, since $g(3)=4\pi a/m$ from Eq. (\ref{24}), the  mean-field energy density is
\be
\cale_0^{(3)} = \frac{2\pi a n^2}{m}.
\label{89}
\ee
In 2D, since $g(2)=4\pi /m\ln(1/\gamma)$ from Eq. (\ref{25}),
the  mean-field energy density is
\be
\cale_0^{(2)} = \frac{2\pi a n^2}{m\ln(1/\gamma)}.
\label{90}
\ee
Therefore, this method  reproduced  the well-known mean-field energy
densities in  Eqs. (\ref{1}) and (\ref{3}) successfully.

%%%%%%%%%%%%%%%%%%%%%%%%%%%%%%%%%%%%%%%%%%%%%
\section{First-order quantum correction}
%in the ground-state energy}
%%%%%%%%%%%%%%%%%%%%%%%%%%%%%%%%%%%%%%%%%%%%%%

The first-order quantum correction of the ground-state energy
is obtained in the following way.
Substituting Eq. (\ref{83}) into  Eqs. (\ref{37}) and (\ref{38}),
we obtain the two variables $X$ and $Y$ at the minimum of $\phi_0$.
\begin{eqnarray}
X&=&\mu - 3 g \phi_0^2 = -2\mu,
\label{91} \\
Y&=&\mu - g \phi_0^2=0.
\label{92}
\end{eqnarray}
The propagator and the dispersion relation
in Eqs. (\ref{eq22}) and (\ref{49}) becomes
\begin{eqnarray}
G(\omega,p)=\dsfrac{1}{\omega^{2}-E_p^2 + i E_p}\left(
\begin{array}{cc}
\frac{p^2}{2m}   & -i \omega \\
\nonumber
 i \omega & \frac{p^2}{2m} + 2\mu
\end{array}\right),\\
\label{93}
\end{eqnarray}
and from Eq. (\ref{87})
\be
E_p=  \sqrt{\frac{p^2}{2m} \left(\frac{p^2}{2m} + 2 \mu \right)}.
\label{94}
\ee
These are the original Bogoliubov results. $E_p$ is gapless
and is linear for small wave-vectors.
For large wave-vectors, the dispersion relation becomes
$E_p \simeq p^2/2m + 2 \mu \simeq p^2/2m + 2 g(D) n.$
Therefore, the $2 g(D) n$ represents the mean-field energy in D-dimensions due to interaction with the condensed particles.

In the Bogoliubov approximation one makes a pair approximation to the Hamiltonian
by neglecting terms with three and four operators \cite{bogoliubov}.
On the other hand, in the Beliaev approximation, one goes one step further by calculating
the leading quantum corrections to the quasi-particle spectrum \cite{beliaev}.
This is done by including all one-loop diagrams.

The free energy can be written with the inverse of the propagator.
From Eq. (\ref{53})
\be
Z={\rm exp}\left\{i\int dt \int d^Dx  \, {\rm det} G^{-1}(\omega,p)\right\}.
\label{50}
\ee
Therefore,
\be
\calf = {\rm Tr} \ln G^{-1}(\omega,p).
\label{51}
\ee
Note that $\ln \rm {det} {\bf A} = {\rm Tr} \ln {\bf A}$ for any matrix ${\bf A}$.
The one-loop contribution to the thermodynamic potential
in D-dimensions, $\Omega_1=i\ln Z$, is obtained by the following way.
\bea
\Omega_1(\mu,\phi)&=& -\frac{i}{2}\int\frac{d\omega}{2\pi}
\int\frac{d^Dp}{(2\pi)^D}\ln \det G(\omega,p)^{-1}
\nonumber \\
&=& \frac{1}{2} \int \frac{d^Dp}{(2\pi)^D} E_p.
\label{96}
\eea
From Eq. (\ref{77}), we have the free energy component
\be
\calf_1(\mu)=\Omega_1(\mu,\phi_0)=\frac{1}{4m} I_{0,-1}(\kappa),
\label{97}
\ee
where $\kappa=4m\mu=4mgn$
and $I_{i,j}$ are  loop integral function in $D$-dimensions, which is defined as \cite{braaten,andersen}
\bea
I_{i,j}^{(D)}(\kappa)
&=& \int\frac{d^Dp}{(2\pi)^D}\frac{p^{2i}}{p^j(p^2+\kappa)^{j/2}},
\label{98}
\eea
where $i$ and $j$ are integers.
This $I_{i,J}$ is expressed by Gamma function, too.
\be
I_{i,j}^{(D)}= \frac{1}{(4 \pi)^{D/2}}
\frac{\Gamma \left( \frac{D+2i-j}{2} \right)
\Gamma \left( -\frac{D+2i-2j}{2} \right)  }
{ \Gamma \left( \frac{D}{2} \right)  \Gamma \left( \frac{j}{2} \right)}
\kappa^{\frac{D+2i-2j}{2}}.
\label{99}
\ee
Note that in D=3, $I_{0,-1}^{(3)}(\kappa) = \kappa^{5/2}/15 \pi^2$,
$I_{1,1}^{(3)}(\kappa)=\kappa^{3/2}/3 \pi^2$,
and $I_{-1,-1}^{(3)}(\kappa)=-\kappa^{3/2}/6 \pi^2$.
It has the useful relation  of the first derivative
\be
\frac{d}{d\kappa} I_{i,j}^{(D)}(\kappa)=-\frac{j}{2}I_{i+1,j+1}^{(D)}(\kappa).
\label{100}
\ee

Ground-state energy density is expressed as
\begin{eqnarray}
\cale_0 + \cale_1 +  \cdots
=\frac{1}{2}g n^2 + \frac{1}{4m} I^{(D)}_{0,-1} (\kappa) + \cdots.
\label{104}
\end{eqnarray}
Therefore, the contribution of the first-order quantum correction in D-dimensions is obtained as
\bea
\frac{\cale_1}{\cale_0}&=&
-\frac{\Gamma\left( -\frac{D+2}{2} \right)\Gamma\left( \frac{D+1}{2} \right)}
{ \pi^{(D+1)/2} \Gamma\left( \frac{D}{2} \right)}\left\{ mg(D) \right\}^{D/2}n^{D/2-1}
\nonumber \\
&=&
-\frac{2^D \pi^{(D^2-2D-2)/4} \Gamma\left( -\frac{D+2}{2}\right)
 \Gamma\left( \frac{D+1}{2}\right)
\gamma^{D/2-1} }
{\Gamma\left( \frac{D}{2}\right) \left\{ 2^{2-D} \Gamma\left(1- \frac{D}{2}\right)\gamma^{D/2-1} + \Gamma\left( \frac{D}{2}-1\right) \right\}^{D/2}}.
\label{105}
\eea
Note that $\gamma=(n_0+n_1+\cdots) a^D$.

%%%%%%%%%%%% FIGURE 2    %%%%%%%%%%%%%%%%%%%%%%%%%%%%%%%
\begin{figure}
\resizebox{!}{0.22\textheight}{\includegraphics{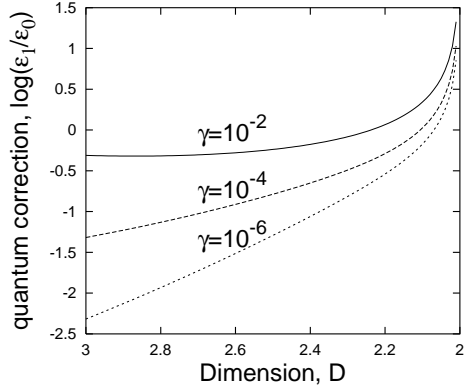}}
\caption{\label{fig:energy}
The first-order quantum correction of the
ground-state energy density in D-dimensions in the logarithmic scale.
The divergence in D=2 is clear.
$\gamma$ itself is a function of the dimensions.}
\end{figure}
%%%%%%%%%%%%%%%%%%%%%%%%%%%%%%%%%%%%%%%%%%%%%%%%%%%%%%%%%%%%

Substituting D=3, we obtain the well-known first-order quantum correction
of the ground-state energy density in 3D as
\be
\frac{\cale_1}{ \cale_0} =\frac{128}{15\sqrt{\pi}} \sqrt{\gamma}.
\label{106}
\ee
Taking the limit of
$D=\lim_{\veps \to 0^+} 2(1+\veps)$, we  obtain the first-order quantum correction
of the ground-state energy density in 2D as
\be
\frac{\cale_1}{\cale_0} \simeq \frac{\gamma^{\veps}}{-\gamma^{\veps} + 1}
\simeq \frac{1}{\veps \ln(1/\gamma)} \gg 1.
\label{108}
\ee
For the singularities of the $\Gamma$ function at non-positive argument,
we used the following expansions
\be
\Gamma(-n+\veps)=\frac{(-1)^n}{n}\left[\frac{1}{\veps} + h(n+1) +\cal{O}(\veps) \right],
\label{109}
\ee
where $h(n+1)=1+1/2+1/3+ \cdots + 1/n -\gamma_E$
and $\gamma_E=0.5772...$ is the Euler constant.
In particular, $\Gamma(\veps)= 1/\veps-\gamma_E$.
Also, we used the approximations:  $\gamma^\veps \simeq 1 + \veps\ln \gamma$.

In 2D there is a divergence, and
we can obtain the two exponents of the divergence
in Eq. (\ref{5}) as $\al=\al'=1$.
We plotted $\cale_1 / \cale_0 $ in FIG. 2 as the functions
of the dimensions and gas parameter in the logarithmic scale.
From the FIG. 2 it is clear that
the expansion in Eq. (\ref{3}) satisfy the perturbative condition of
$\cale_1/\cale_0 \ll 1 $ only when $D > 2.2$.
In low dimensions of  $D < 2.2$, the first-order ground-state energy density is not perturbative regardless of the magnitude of the gas parameter.

%%%%%%%%%%%%%%%%%%%%%%%%%%%%%%%%%%%%%%%%%%%%%%%%%%%%%%%%%%%%%%%%
%\section{first-order number density in the ground-state}
%%%%%%%%%%%%%%%%%%%%%%%%%%%%%%%%%%%%%%%%%%%%%%%%%%%%%%%%%%%%%

The number density is obtained from Eqs. (\ref{27})-(\ref{eq7}) as
\be
n=\phi^2 + \frac{1}{2}\langle \psi_1^2 + \psi_2^2 \rangle.
\label{130}
\ee
Therefore, taking the one-loop effects into account
\bea
n_0 + n_1
&=& n_0 + \frac{i}{2}\int\frac{d\omega}{2\pi}\frac{d^Dp}{(2\pi)^D}
\left[ \frac{p^2/2m + 2m E_p^2/p^2}{\omega^2 -E_p^2 + i E_p} \right]
\nonumber \\
&=& n_0 + \frac{1}{4}I_{1,1}(4mgn) + \frac{1}{4}I_{-1,-1}(4mgn)
\nonumber \\
 & =& n_0 \left[ 1+ \frac{ \Gamma\left(-\frac{D}{2}\right)
 \left\{ 2\Gamma\left( \frac{D+1}{2}\right) - \Gamma\left( \frac{D-1}{2}\right) \right\}  }
 {  2^{3}\pi^{(D+2)/2}\Gamma\left(\frac{D}{2}\right)  }(mg)^{D/2} n_0^{(D-2)/2} \right] .
  \label{134}
  \eea

  Then, the first-order quantum correction of the number density in D-dimensions
  is obtained as
 \bea
\frac{n_1}{n_0}
 &=&\frac{\pi^{(D^2-2D-2)/4}}{2^{3-D}}
   \frac{  \Gamma\left(-\frac{D}{2}\right)
    \left\{ 2\Gamma\left( \frac{D+1}{2}\right) - \Gamma\left( \frac{D-1}{2}\right) \right\} }
   {   \Gamma\left(\frac{D}{2}\right)
   \left\{ 2^{2-D}\Gamma\left( 1-\frac{D}{2} \right) \gamma^{D/2-1} + \Gamma\left( \frac{D}{2} -1\right)\right\}^{D/2}} \gamma^{D/2-1}.
     \label{135}
 \eea
In 3D we obtain the well-known first-order quantum depletion as
\be
\frac{n_1}{n_0} = \frac{8}{3\sqrt{\pi}} \sqrt{\gamma}.
\label{138}
\ee
In 2D taking the limit of
$D=\lim_{\veps \to 0^+} 2(1+\veps)$,
we  obtain the first-order quantum depletion as
\bea
\frac{ n_1 } { n_0 }
 &\simeq&
 \frac{ \gamma^\veps }
 {  -\frac{1}{\veps}\gamma^{\veps} + \frac{1}{\veps} }.
 \nonumber \\
 &\simeq& \frac{1}{\ln(1/\gamma)}.
\label{156}
\eea
We used the approximations:  $\Gamma(-\frac{D}{2}) \sim 1/\veps$ and $2\Gamma(\frac{D+1}{2}) -\Gamma(\frac{D-1}{2}) \sim 2\sqrt{\pi}\veps$ .
This result is the same as the Schick's value in Eq. (\ref{4}).
We plotted $ n_1 / n_0 $ as a function of dimensions and gas parameter in Fig. 3.
 It is increasing monotonically as the dimension is decreasing from 3 to 2.
The depletion is stronger in low dimensions.

%%%%%%%%%%%% FIGURE 3    %%%%%%%%%%%%%%%%%%%%%%%%%%%%%%%
\begin{figure}
\resizebox{!}{0.22\textheight}{\includegraphics{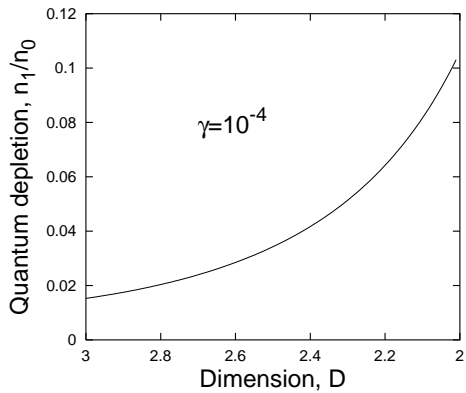}}
%\resizebox{!}{0.2\textheight}{\includegraphics{gas4c.eps}}
\caption{\label{fig:number}
The first-order quantum correction of the number density
at $\gamma=10^{-4}$ as a function the dimensions.
}
\end{figure}
%%%%%%%%%%%%%%%%%%%%%%%%%%%%%%%%%%%%%%%%%%%%%%%%%%%%%%%%%%%%

%%%%%%%%%%%%%6666666666666%%%%%%%%%%%%%%%%%%%%%
\section{Summary }
%%%%%%%%%%%%%6666666666666%%%%%%%%%%%%%%%%%%%%%%
EFT is a theory of symmetries.
Once the symmetries have been identified, one writes down
the most general local effective Lagrangian consistent with these symmetries.
At zero temperature, the symmetries are Galilean invariance, time-reversal symmetries,
and the global phase symmetry.
These symmetries restricts the possible terms in the effective action.

The first-order quantum correction to the ground-state energy density of
a 2D uniform Bose atoms has the form of the divergence:
$\cale_1/\cale_0 \sim |D-2|^{-\al} |\ln\gamma|^{-\al'}$.
Applying an effective field theory method to the hard-sphere boson system
in D-dimensions,
we obtained the D-dimensional free energy and thermodynamic potential
up to the one-loop result.
This general form in D-dimensions reproduced
every well-known result  in 3D and 2D.

Then, taking the limit of $D=\lim_{\veps \to 0^+} 2(1+\veps)$,
 we obtained  the zeroth- and first-order quantum correction in 2D.
Finally, we obtained the two exponents of divergence $\al$ and $\al'$ analytically
as  $\alpha=\alpha'=1$.
We also showed that the perturbative expression of the ground-state energy density
is effective only in low dimensions of $D < 2.2$.

%%%%%%%%%%%%%%%%%%%%%%%%%%%%%%%%%%%%%%%%%%%%%%%%%%%%%%%%%%%%
\begin{acknowledgments}
This work is supported by Korea Science and Engineering
Foundation(KOSEF) and Australian Academy of Science(AAS) scientific
exchange program in 2008.
\end{acknowledgments}

%%%%%%%%%%%%%%%%%%%%%%%%%%%%%%%%%%%%%%%%%%%%%%
\bb{99}
%%%%%%%%%%%%%%%%%%%%%%%%%%%%%%%%%%%%%%%%%%%%%%
\bi{bogoliubov} N. N. Bogoliubov, J. Phys. (U.S.S.R.) {\bf 11}, pp. 23-32 (1947).
\bi{lee} T. D. Lee and C. N. Yang, Phys. Rev. {\bf 105}, pp. 1119-1120 (1957).
\bi{brueckner} K. A. Brueckner, K. Sawada. Phys. Rev. {\bf 106}, pp. 1117-1127 (1957).
\bi{beliaev} S. T. Beliaev, Sov. J. Phys. {\bf 7}, pp. 289-299 (1958).
\bi{wu} T. T. Wu, Phys. Rev. {\bf 115}, pp. 1390-1404 (1959).
\bi{hugenholtz} N.M. Hugenholtz and D. Pines, Phys. Rev. {\bf 116}, pp. 489-506 (1959).
\bi{sawada} K. Sawada, Phys. Rev. {\bf 116}, 1344-1358 (1959).
\bi{braaten} E. Braaten and A. Nieto, Euro. Phys. J. B {\bf 11}, pp. 143-159 (1999).
\bi{georgi} H. Georgi, Ann. Rev. Nucl. Part. Sci. {\bf 43}, pp. 209-252 (1993).
\bi{kaplan} D. B. Kaplan, nucl-th/9506035 (1995).
\bi{manohar} A. V. Manohar, {\it Effective field theories,} in {\it Perturbative
    and Nonperturbative Aspects of Quantum Field Theory},
    Ed. by H. Latal and W. Schweiger  (Springer-Verlag, 1997).
\bi{ketterle} A. G\"{o}rlitz, J. M. Vogels, A. E. Leanhardt, C. Raman,
T. L. Gustavson, J. R. Abo-Shaeer, A. P. Chikkatur, S. Gupta, S. Inouye,
T. Rosenband, and W. Ketterle, \prl {\bf 87}, pp. 130402:1-4 (2001).
\bi{schick} M. Schick, \pra, {\bf 3}, pp. 1067-1073 (1971).
\bi{andersen} J. O. Andersen, \rmp {\bf 76}, pp. 599-639 (2004).
\bi{rakimov} A. Rakhimov, C. K. Kim, S.-H. Kim, and J. H. Yee, \pra {\bf 77},
    pp. 033626:1-9 (2008).
\bi{nogueira} F. S. Nogueira and H. Kleinert, \prb {\bf 73}, pp. 104515:1-9 (2006).
There is a factor of 2 difference in the exponent of $\gamma$ of $g(D)$ in Eq. (\ref{23}).
We  took $pa \rightarrow \sqrt{na^2} $ instead Nogueira's Choice of
 $pa \rightarrow na^2 $, where $p$ is the momentum.
\bi{kim}  S.-H. Kim, C. Won, S. D. Oh, and W. Jhe,  cond-mat/9904087 (1999).
\bi{lieb} E. H. Lieb, R. Seiringer, and J. Yngvason,
Commun. Math. Phys. {\bf 224}, pp. 17-31 (2001).
%%%%%%%%%%%%%%%%%%%%%%%%%%%%%%%%%%%%%%%%%%%
\eb
%%%%%%%%%%%%%%%%%%%%%%%%%%%%%%%%%%%%%%%%%%
\edc